\documentstyle[12pt]{article}
\begin{document}
\begin{center}

{\Large \bf Adelic strings and noncommutativity }

\bigskip

{\large Branko Dragovich }

{\it Institute of Physics,
P.O.Box 57, 11001 Belgrade, Yugoslavia }

{\it Steklov Mathematical Institute, Gubkin St. 8, 117966 Moscow,
Russia}

\end{center}

\begin{abstract}
We consider adelic approach to strings and spatial noncommutativity.
Path integral method to string amplitudes is emphasized. Uncertainties
in spatial measurements in quantum gravity are related
to noncommutativity between coordinates. $p$-Adic
and adelic Moyal products are introduced. In particular, $p$-adic
and adelic counterparts of some real noncommutative scalar solitons
are constructed.
\end{abstract}

\date{}

\section{Introduction}

There is a common belief that an appropriate description of the
Planck scale phenomena needs some new physical principles and
nonstandard mathematical methods. To this end, we consider adelic
approach to string theory and spatial noncommutativity.

The space of adeles ${\bf A}$ is a mathematical instrument, which
unifies real numbers (with archimedean geometry) and all $p$-adic numbers
(with their nonarchimedean geometries). Note that all numerical
experimental data belong to the field of rational numbers ${\bf Q}$, and
that ${\bf Q}$ is a dense subfield of the field of real numbers ${\bf R}$
and the fields of $p$-adic numbers ${\bf Q}_p$ ($p$ denotes a prime number,
i.e. $p = 2, 3, 5,\cdots$). There is a sense to expect that basic mathematical
methods and fundamental physical laws are invariant under interchange
of ${\bf R}$ and ${\bf Q}_p$ \cite{drag-vol0},
and such invariance  has a place in adelic formalism.
Possible spatial $p$-adic effects become sensitive in the vicinity
of the characteristic length of a quantum system. Since the Planck length
\begin{equation}
\label{drag-eq1}
   \ell_0 =\sqrt\frac{\hbar G}{c^3} \sim 10^{-33} cm
\end{equation}
is the natural one for quantum gravity, the Planck scale physics should
exhibit some $p$-adic effects. Since 1987, there have been interesting
investigations of $p$-adic and adelic string models
(for an early review, see \cite{drag-fre1,drag-vol1}).
As a basis for  systematic investigations, $p$-adic
\cite{drag-vol2} and adelic \cite{drag-dra1} quantum mechanics
are formulated.

It is well-known that an application of quantum-mechanical principles
to general relativity leads to the uncertainties in measurements of
very short distances in the form
\begin{equation}
\label{drag-eq2}
\Delta x^i \Delta x^j \geq \ell_0^2, \ \ \ i,j =1,2,\cdots,n,
\end{equation}
where $n$ is spatial dimensionality. This fact requires a reconsideration
of many our usual notions about the spacetime structure approaching to the
Planck scale.  The uncertainty (\ref{drag-eq2}) has to be a consequence of
the corresponding noncommutativity between operators of space coordinates
in the Hilbert space.
This conclusion is an
analogue of the similar situation in ordinary quantum mechanics:
the uncertainty
 $\Delta x\Delta k\geq \frac{\hbar}{2}$
is a direct consequence of the noncommutativity in the form of the
Heisenberg algebra $[\hat x,\hat k]=i\hbar$, where
$x$ and $k$ are coordinates of the phase space. Thus, we see that
the uncertainty (\ref{drag-eq2}) implies
noncommutative geometry given by the commutation relation
\begin{equation}
\label{drag-eq3}
[\hat x^{i} , \hat x^j] = i\hbar \theta^{ij}, \ \  \theta^{ij} = -\theta^{ji},
\end{equation}
where $\theta^{ij}=\theta \varepsilon^{ij} \ $ ($\varepsilon^{ij} = 1$ if
$i< j$) and $\theta = 2\ell_0^2/\hbar$.

If the uncertainty
(\ref{drag-eq2}) holds also for the case $i=j$, i.e. $\Delta x^i \geq\ell_0$,
then it leads to the direct
restriction on application of real numbers below $\ell_0$
and gives rise to employment of $p$-adic numbers. However, it is not clear
how this kind of uncertainty can be related to some spatial noncommutativity,
and consequently we will omit here further discussion of this subject.

One of the very interesting and fruitful recent developments in string
theory (for a reviev, see \cite{drag-schw1,drag-sen1}) has been
noncommutative geometry and the corresponding noncommutative field theory.
This subject started to be very actual after Connes, Douglas and Schwarz
shown \cite{drag-con1} that gauge theory on noncommutative torus describes
compactifications of M-theory to tori with constant background three-form
field. Noncommutative field theory may be regarded
as a deformation of the ordinary one in which field multiplication is
replaced by the Moyal (star) product
\begin{equation}
\label{drag-eq4}
(f\star g)(x) =
\exp\left[\frac{i \hbar}{2} \theta^{ij}\frac{ \partial}{\partial y^{i}}
\frac{ \partial}{\partial z^j}\right]
f(y)g(z)\vert_{y=z =x},
\end{equation}
where $x^1,x^2,\cdots,x^d$ denote  coordinates of noncommutative space, and
$\theta^{ij} =-\theta^{ji}$ are noncommutativity parameters.
There are many properties of D-brane dynamics which may be studied  by
noncommutative field theory. In particular, it enables to investigate a
mixing of the UV and  IR effects, and the tachyon condensation.
Replacing the ordinary product
between coordinates by the Moyal product (\ref{drag-eq4}), we have
\begin{equation}
\label{drag-eq5}
 x^{i}\star x^j - x^j \star x^{i} = i\hbar \theta^{ij} ,
\end{equation}
which resembles the usual Heisenberg algebra. Comparing the above equations
it follows that the algebra (\ref{drag-eq3}) can be realized by ordinary
coordinates using the Moyal product between them.

In the Section 2 we provide reader with some very basic facts on
$p$-adic numbers and adeles.
Section 3 is devoted to  adelic strings.
In the last section we consider some $p$-adic and adelic aspects
of spatial noncommutativity.

\section{$p$-Adic numbers and adeles}

In order to introduce $p$-adic numbers it is suitable to start
from ${\bf Q}$, since ${\bf Q}$ is the simplest field of numbers of
characteristic $0$ and it contains results of all physical measurements.
Any non-zero rational number can be presented as infinite expansions
into the two quite different forms. The usual one is to the base $10$,
{i.e.}
\begin{equation}
\label{drag-eq6}
\sum_{k=n}^{-\infty} a_k 10^k ,    \quad a_k =0,\cdots, 9 ,
\end{equation}
and the other one is to the base $p$ ($p$ is a prime number) and reads
\begin{equation}
\label{drag-eq7}
\sum_{k=m}^{+\infty} b_k p^k ,   \quad b_k = 0,\cdots, p-1 ,
\end{equation}
where $n$ and $m$ are some integers. These representations have the usual
repetition of digits, but expansions are in the mutually opposite directions.
The series (\ref{drag-eq6})
and (\ref{drag-eq7}) are convergent with respect to the usual absolute value
$| \cdot |_\infty $  and
$p$-adic absolute value $| \cdot |_p$, respectively. Allowing arbitrary
distributions of digits, we obtain standard
representation of real numbers (\ref{drag-eq6}) and $p$-adic numbers
(\ref{drag-eq7}).
${\bf R}$ and ${\bf Q}_p$ exhaust all number fields which contain ${\bf Q}$
as a dense subfield.
They have many distinct geometric and algebraic properties. Geometry of
$p$-adic numbers is the nonarchimedean one.

There are mainly two kinds of analysis on ${\bf Q}_p$ based on two
different mappings: ${\bf Q}_p\to {\bf Q}_p$ and ${\bf Q}_p \to {\bf C} $.
We use here both of them,  in classical and quantum $p$-adic models,
respectively.

Elementary $p$-adic functions are given by the same series  as in
the real case, but their regions of convergence are usually different.
For instance, $\exp x = \sum_{n=0}^\infty {{x^n}\over{n!}}$
and $\ln x = \sum_{n=1}^\infty (-1)^{n+1} {{(x-1)^n}\over{n}} $
converge if $| x |_p < | 2 |_p$ and $|
x-1 |_p <1$, respectively. Derivatives of $p$-adic valued
functions are also defined as in the real case, but using $p$-adic
norm instead of the absolute value. As a definite $p$-adic valued integral
we take difference of the corresponding antiderivative in end points.

Usual complex-valued $p$-adic functions are:  ({\it i}) an additive
character $\chi_p (x) =\exp 2\pi i\{x\}_p$, where $\{x\}_p$ is
the fractional part of $x\in {\bf Q}_p$,  ({\it ii}) a multiplicative
character $\pi_s (x) = \vert x\vert_p^s$, where
$s\in {\bf C}$, and  ({\it iii})  locally constant functions with
compact support, like, e.g. $\Omega (| x |_p) = 1$ if
$| x |_p \leq 1$ and $\Omega (| x |_p) = 0$ otherwise.
There is well defined Haar measure and integration. For instance,
\begin{equation}
\label{drag-eq8}
 \int_{{\bf Q}_p}\chi_p (\alpha x^2 + \beta x) dx = \lambda_p (\alpha)
| 2\alpha |_p^{-\frac{1}{2}} \chi_p \left(
-\frac{\beta^2}{4\alpha} \right), \quad \alpha \neq 0 ,
\end{equation}
where $\lambda_p(\alpha)$ is an arithmetic function \cite{drag-vol1}.
For much more on $p$-adic numbers and $p$-adic
analysis one can see, e.g. \cite{drag-vol1,drag-gel1,drag-sch1}.

An adele  $x_{{\bf A}}$  \cite{drag-gel1} is an infinite sequence
\begin{equation}
\label{drag-eq9}
   x_{{\bf A}} = (x_\infty , x_2 , \cdots, x_p , \cdots) ,
\end{equation}
where $x_\infty\in {\bf R}$ and $x_p\in {\bf Q}_p$ with the restriction that
for all but a finite
set ${\bf S}$ of primes $p$ we have $x_p\in {\bf Z}_p$ .
Here ${\bf Z}_p = \{x\in {\bf Q}_p : \vert x\vert_p \leq 1     \}  $
is the ring of $p$-adic integers.
All adeles make a ring with respect to componentwise addition and
multiplication. It is convenient to present the ring of adeles
${\bf A}$  in the  form
\begin{equation}
\label{drag-eq10}
  {\bf A} = \cup_S {\cal A}(S) ,   \ \ {\cal A}(S) =
{\bf R}\times \prod_{p\in S} {\bf Q}_p   \times \prod_{p\not\in S} {\bf Z}_p.
\end{equation}
${\bf A}$ is also locally compact
topological space with the Haar measure. There are two kinds of
analysis over ${\bf A}$, which generalize the corresponding analyses
over ${\bf R}$ and ${\bf Q}_p$.

\section{ Adelic strings }

A notion of $p$-adic string was introduced by Volovich in \cite{drag-vol3},
where the hypothesis on the existence of nonarchimedean geometry
at the Planck scale was made, and string theory with $p$-adic
numbers was initiated. In particular, generalization of the usual
Veneziano and Virasoro-Shapiro amplitudes with complex-valued
multiplicative characters over various number fields was proposed
and  $p$-adic valued Veneziano amplitude was constructed by means
of $p$-adic interpolation. Very successful  $p$-adic analogues of
the Veneziano and Virasoro-Shapiro amplitudes were proposed in
\cite{drag-fre2} as the corresponding  Gel'fand-Graev \cite{drag-gel1}
beta functions. Using this approach, Freund and Witten
obtained \cite{drag-fre3} an attractive adelic formula, which states
that the product of the crossing symmetric Veneziano (or Virasoro-Shapiro)
amplitude and its all $p$-adic counterparts equals unit (or a definite
constant). This gives possibility to consider an ordinary four-point
function, which is rather complicate, as an infinite product of its
inverse $p$-adic analogues, which have simple forms. These first papers
induced an interest in various aspects of $p$-adic string theory
(for a review, see \cite{drag-fre1,drag-vol1}). A recent interest in
$p$-adic string theory has been mainly related to the generalized
adelic formulas for four-point string amplitudes \cite{drag-vla1},
the tachyon condensation \cite{drag-sen2}, and the new  promising
adelic approach \cite{drag-dra2}.

Like in the ordinary string theory, the starting point
of $p$-adic  strings is a construction of the corresponding
scattering amplitudes.  Recall that the ordinary crossing symmetric
Veneziano amplitude
can be presented in the following forms:
\begin{equation}
\label{drag-eq11}
  A_\infty(a,b)= g^2 \int_{\bf R} | x |_\infty^{a-1} 
| 1-x |_\infty^{b-1} dx
\end{equation}
\begin{equation}
\label{drag-eq12}
 = g^2 \left[\frac{\Gamma{(a)} \Gamma{(b)} }{\Gamma{(a+b)}} +
 \frac{\Gamma{(b)}\Gamma{(c)}}{\Gamma{(b+c)}} +
 \frac{\Gamma{(c)}\Gamma{(a)}}{\Gamma{(c+a)}}\right]
\end{equation}
\begin{equation}
\label{drag-eq13}
 = g^2 \frac{\zeta(1-a)}{\zeta(a)} \frac{\zeta(1-b)}{\zeta(b)}
 \frac{\zeta(1-c)}{\zeta(c)}
\end{equation}
\begin{equation}
\label{drag-eq14}
=g^2 \int {\cal D}X \exp\left(  -\frac{i}{2\pi}
 \int d^2 \sigma \partial^\alpha X_\mu \partial_\alpha X^\mu \right)
 \prod_{j=1}^4 \int d^2 \sigma_j \exp\left( i k_\mu^{(j)} X^\mu
\right) ,
\end{equation}
where $\hbar=1,\ T=1/\pi$, and $a=-\alpha (s) = - 1 -\frac{s}{2},
\ b=-\alpha (t), \ c=-\alpha (u)$ with the condition
$s+t+u = -8$, i.e. $a+b+c=1$.

To introduce the corresponding $p$-adic Veneziano amplitude there
is a sense to consider $p$-adic analogues of all the above four
expressions. $p$-Adic generalization of the first expression was
proposed in \cite{drag-fre2} and it reads
\begin{equation}
\label{drag-eq15}
 A_p (a,b) = g_p^2 \int_{{\bf Q}_p} \vert x\vert_p^{a-1}
 | 1-x |_p^{b-1} dx ,
\end{equation}
where $| \cdot |_p$ denotes $p$-adic absolute value.
 In this case only
string world-sheet parameter $x$ is treated as $p$-adic variable,
and all other quantities have their usual (real) valuation.
An attractive adelic formula of the form
\begin{equation}
\label{drag-eq16}
 A_\infty (a,b) \prod_p A_p (a,b) =1
\end{equation}
was found \cite{drag-fre3}, where $A_\infty (a,b)$ denotes the usual
Veneziano amplitude (\ref{drag-eq11}). A similar product formula holds
also for the
Virasoro-Shapiro amplitude. These  infinite products are divergent,
but they can be successfully regularized.
Unfortunately, there is a problem to extend
this formula to the higher-point functions.

$p$-Adic analogues of (\ref{drag-eq12}) and (\ref{drag-eq13})
were also proposed in \cite{drag-vol3}
and \cite{drag-dra3}, respectively. In these cases,
world-sheet, string momenta and amplitudes are manifestly $p$-adic.
Since these string amplitudes are $p$-adic valued functions, it is not
enough clear their physical meaning.

Expression (\ref{drag-eq14}) is based on Feynman's functional integral method,
which is a useful tool in all quantum theory and has successful $p$-adic
generalization \cite{drag-dra4}.
Its $p$-adic counterpart,
proposed in \cite{drag-dra2}, has been elaborated  \cite{drag-dra5} and
deserves further study. Note that in this approach,  $p$-adic
string amplitude is complex-valued, while not only the world-sheet
parameters but also target space coordinates and string momenta
are $p$-adic variables. Such $p$-adic generalization is a natural extension
of the formalism of $p$-adic \cite{drag-vol2} and adelic \cite{drag-dra1}
quantum mechanics to string theory.

Our starting point to unified path integral approach to ordinary and
$p$-adic $N$-point bosonic string amplitudes at the tree level is
\begin{equation}
\label{drag-eq17}
A_v(k_1,\cdots,k_N) = g_v^{N-2} \prod_{j=1}^N \int d^2\sigma_j
\int  \chi_v \left( -\frac{1}{h}
\int {\cal L}(X^\mu,\partial_\alpha X^\mu) d^2\sigma \right) {\cal D}_v X,
\end{equation}
where $v=\infty, 2,\cdots, p, \cdots, \ \ \mu = 0,1,\cdots,25,
\ \ \alpha = 0,1$, and $\chi_\infty(a) = \exp(-2\pi i a), \
\chi_p(a) = \exp(2\pi i \{ a\}_p)$.
The above Lagrangian is
\begin{equation}
\label{drag-eq18}
{\cal L} = -\frac{T}{2} \partial_\alpha X^\mu (\sigma,\tau)
\partial^\alpha X_\mu (\sigma,\tau) + \sqrt{-1} \sum_{j=1}^N
k_\mu^{(j)}X^\mu (\sigma,\tau) \delta (\sigma -\sigma_j) \delta
(\tau - \tau_j).
\end{equation}

In fact, our approach is adelic and based on the following assumptions:
({\it i}) spacetime and matter are adelic at the Planck (M-theory)
scale, ({\it ii}) Feynman's path integral method is an inherent ingredient
of quantum theory, and ({\it iii}) adelic quantum theory is a more complete
theory than the ordinary one. Consequently, a string is an adelic object
which has simultaneously real and all $p$-adic characteristics. Here the term
$p$-adic  (real) string is related to string with dominant $p$-adic (real)
properties. The target space and world-sheet are adelic spaces. Adelic
Feynman's path integral is an infinite product of the ordinary one and all
$p$-adic counterparts \cite{drag-dra6}.

The corresponding adelic string amplitude is
$$
 A (k_{\bf A}^{(1)},\cdots, k_{\bf A}^{(N)})
$$
\begin{equation}
\label{drag-eq19}
= A_\infty (k_\infty^{(1)},\cdots,k_\infty^{(N)})
 \prod_{p\in{\bf S}} A_p (k_p^{(1)},\cdots,k_p^{(N)})
 \prod_{p\not\in {\bf S}} A_p (k_p^{(1)},\cdots,k_p^{(N)}) ,
\end{equation}
where $k_{\bf A}^{(i)}$ is an adele, i.e.
\begin{equation}
\label{drag-eq20}
k_{\bf A}^{(i)} = (k_\infty^{(i)}, k_2^{(i)}, \cdots, k_p^{(i)},\cdots)
\end{equation}
with the restriction that $k_p^{(i)} \in {\bf Z}_p$ for all
but a finite set ${\bf S}$
of primes $p$. The topological ring of adeles $k_{\bf A}^{(i)}$
provides a framework for simultaneous and unified consideration
of real and $p$-adic string momenta.
Adelic string amplitude contains nontrivial $p$-adic modification
of the ordinary one. An evaluation of the above $p$-adic and adelic
amplitudes will be presented in detail elsewhere \cite{drag-dra5}.

\section{$p$-Adic  and adelic noncommutativity}

There is a noncommutative scalar soliton \cite{drag-str1}
\begin{equation}
\label{drag-eq21}
   \phi (x^1, x^2) = 2 \exp{\left(-\frac{(x^1)^2 + (x^2)^2}{\theta} \right)}
\end{equation}
which is the simplest nontrivial solution (trivial solutions are $\phi =0$
and $\phi = 1$) of the equation
\begin{equation}
\label{drag-eq22}
(\phi \star \phi) (x) = \phi (x),
\end{equation}
where $\star$ denotes the Moyal product (\ref{drag-eq4}) with
$\theta^{ij} = \theta \varepsilon^{ij}/\hbar$.
The solution (\ref{drag-eq21}) of the equation (\ref{drag-eq22})
extremises energy   in noncommutative scalar field theory
\cite{drag-str1} with the potential
\begin{equation}
\label{drag-eq23}
V(\phi) = \frac{1}{2}m^2\phi\star\phi - \frac{1}{3}\phi\star\phi\star\phi ,
\end{equation}
 where $m=1$ and  the kinetic term is neglected in the limit
$\theta \rightarrow \infty$.
An intriguing similarity between the above noncommutative scalar soliton and
a solitonic brane solution \cite{drag-sen2} in an effective $p$-adic string
theory \cite{drag-fra1} is discussed in \cite{drag-dra7}.

 This two-dimensional noncommutative scalar field model can be extended
to the more general case with
 \begin{equation}
 \label{drag-eq24}
 V(\phi) = \frac{1}{2} m^2 \phi_\star^2 - \frac{c_{k+1}}{k+1}
\phi_\star^{k+1} ,
 \end{equation}
 where $\phi_\star$ denotes that fields are multiplied by the star product, and
 $\phi \equiv \phi (x^1,\cdots,x^n)$ with even $n$  spatial directions.
The corresponding equation
\begin{equation}
\label{drag-eq25}
c_{k+1} \phi_\star^k(x) = m^2 \phi(x)
\end{equation}
has the solution
\begin{equation}
\label{drag-eq26}
\phi (x) = 2^{\frac{n}{2}} \left( \frac{m^2}{c_{k+1}} \right)^{\frac{1}{k}}
\exp{\left( -\frac{1}{\theta} \sum_{i=1}^n (x^i)^2\right)}.
\end{equation}

The above formulas  (\ref{drag-eq21}) - (\ref{drag-eq26})
are related to the case with real numbers. Now we introduce their
$p$-adic and adelic generalization. Let the Moyal product for $p$-adic
valued functions $f$ and $g$ be
\begin{equation}
\label{drag-eq27}
(f\star g)(x) =
\exp\left[\frac{\sqrt{-1} }{2} \theta\varepsilon^{ij}
\frac{ \partial}{\partial y^{i}}\frac{ \partial}{\partial z^j}\right]
f(y)g(z)\vert_{y=z =x},
\end{equation}
where $x,y,z\in {\bf Q}_p$. Then the equations (\ref{drag-eq21}) -
(\ref{drag-eq26})
also hold in the $p$-adic case. The region of convergence of exponential
functions is given by inequality $|x^i|_p < |2\theta|_p^{1/2}$.
Thus, the equations (\ref{drag-eq22}) and (\ref{drag-eq25}) have real
$\phi(x_\infty)$ and $p$-adic $\phi(x_p)$ solutions of the same
functional form $\phi$, i.e. they are invariant under interchange
of ${\bf R}$ and ${\bf Q}_p$. Moreover, they have natural adelic solutions
\begin{equation}
\label{drag-eq28}
\phi_{\bf A}(x_{\bf A}) = (\phi(x_\infty), \phi(x_2), \cdots,
\phi(x_p), \cdots)
\end{equation}
where parameters $m,\theta,c_{k+1} \in {\bf Q}$. Hence, one can
consider not only real, but also $p$-adic and adelic
noncommutative scalar solitons.
This subject will be presented in more details in \cite{drag-dra8}.

It is worth noting that one can introduce \cite{drag-dra9} the Moyal product
in complex-valued $p$-adic quantum mechanics and it reads
\begin{equation}
\label{drag-eq29}
(f \ast g)(x)=\int_{{\bf Q}_p^n}\int_{{\bf Q}_p^n} dk dk'
\ \chi_p(-\frac{1}{h}(x^ik_i+x^jk'_j)+\frac{1}{2h} k_ik'_j\theta^{ij})\tilde
f(k)\tilde g(k'),
\end{equation}
where $n$ denotes  spatial dimensionality. This is a direct $p$-adic
analogue of the integral form of the Moyal product in the real case.
The corresponding adelic version of (\ref{drag-eq29}) is
\begin{equation}
\label{drag-eq30}
(f\ast g)_{\bf A} (x_{\bf A}) = (f\ast g)_\infty (x_\infty)
\prod_p (f\ast g)_p (x_p)
\end{equation}
\begin{equation}
\label{drag-eq31}
= \prod_v \int_{{\bf Q}_v^n}\int_{{\bf Q}_v^n} dk_v dk'_v
\ \chi_v(-\frac{1}{h}(x_v^ik_{vi} +x_v^jk'_{vj})+
\frac{1}{2h} k_{vi}k'_{vj}\theta_v^{ij})\tilde
f_v(k_v)\tilde g_v(k'_v),
\end{equation}
where $x_p^i \in {\bf Z}_p$ and $\theta_p^{ij} \in {\bf Z}_p$
for almost all primes $p$.

\bigskip

\noindent
{\bf Acknowledgments}

\noindent
Author wishes to thank organizers of the XXXVII Karpacz Winter School
of Theoretical Physics: New Developments in Fundamental Interaction
Theories, (February 6-15, 2001), for their invitation to participate and give
a talk. The work on this paper was supported in part by RFFI grant
990100866.

\end{document}